\documentclass[fleqn,usenatbib]{mnras}

\usepackage{newtxtext,newtxmath}

\usepackage[T1]{fontenc}

\DeclareRobustCommand{\VAN}[3]{#2}
\let\VANthebibliography\thebibliography
\def\thebibliography{\DeclareRobustCommand{\VAN}[3]{##3}\VANthebibliography}

\usepackage{graphicx}	
\usepackage{amsmath}	
\usepackage{subcaption}
\usepackage{float}

\title[Triggered Fragmentation]{Triggered Fragmentation in Self-Gravitating Protoplanetary discs: Cooling, Mass Movement and Instability}

\author[P. Rawat et al.]{
Pratishtha Rawat,$^{1,2}$\thanks{E-mail: Pratishtha.Rawat@warwick.ac.uk}
Farzana Meru,$^{1, 2}$
Rebecca Nealon$^{1, 2, 3}$
\\
$^{1}$Centre for Exoplanets and Habitability, University of Warwick, Coventry CV4 7AL, UK\\
$^{2}$Department of Physics, University of Warwick, Coventry CV4 7AL, UK\\
$^{3}$School of Physics and Astronomy, Monash University, Clayton, Victoria, 3800, Australia\\
}

\date{Accepted 2026 August 12. Received 2026 August 3; in original form 2025 December 16}

\pubyear{\the\year{}}

\begin{document}
\label{firstpage}
\pagerange{\pageref{firstpage}--\pageref{lastpage}}
\maketitle

\begin{abstract}
Previous three-dimensional hydrodynamical simulations of gravitationally unstable discs have shown that the formation of a single fragment can trigger the formation of subsequent fragments. This behaviour was attributed to changes in the surface mass density caused by the interaction between the first fragment and the disc material. This study reanalyses those simulations to see if the surface mass density was the sole driver. Our results reveal that both surface mass density and sound speed can contribute to the formation of additional fragments. Beyond the general cooling typical of such discs, the inwards movement of cool material from the outer disc, driven by the formation of the first fragment, can enhance fragmentation in the inner regions. We also identify that interactions between the midplane gas and the cooler upper layers of the disc can facilitate additional cooling. Triggered fragmentation can create a unique planet-formation environment by redistributing material across the disc, potentially leading to chemically distinct fragments from those formed in-situ.

\end{abstract}

\begin{keywords}
protoplanetary discs -- hydrodynamics -- gravitation -- instabilities -- accretion, accretion discs -- planets and satellites: formation
\end{keywords}



\section{Introduction}\label{intro}

The spatial distribution and physical characteristics of protoplanetary discs offer insights into the underlying mechanisms governing planet formation. Observations have identified several types of structures within these discs, including rings, gaps, cavities, spirals, warps, and fragments (\citealt{alma2015}; \citealt{andrews2018}). The complexity and diversity of these features fundamentally influence every aspect of planet formation and help explain the wide range of planetary systems observed. Recent discoveries of such structures at early stages of disc evolution in systems like in IRS 63 \citep{cox2020, cristian2023}, L1448 IRS3B \citep{tobin2016}, GY 91 \citep{sheehan2018}, and Elias 2-27 \citep{perez2016}, suggest that planet formation may begin earlier than predicted by standard core accretion models. This has renewed interest in gravitational instability (GI), which can operate efficiently in the massive, young discs typical of early star formation phases.\\

Young protoplanetary discs undergo a self-gravitating phase, during which the disc’s own gravity plays a major role in its evolution and can lead to GI and fragmentation (\citealt{boss2000, rice2004}; see reviews by \citealt{helled2014, kratter2016}). Studies show that disc-to-star mass ratios, $M_{\rm disc}/M_\star$, values of 0.1–1 are common in these early Class 0/I stages \citep{bate2018, manara2023} when substantial mass accretes onto the disc, placing such systems in precisely the regime where self-gravity can shape their evolution. This regime is particularly important because it marks a stage where the disc's internal dynamics are not dominated solely by the central protostar, potentially leading to the formation of bound structures such as stars, brown dwarfs and giant planets (\citealt{may2004}; \citealt{nayak2007}; \citealt{stam2007}). Although self-gravity may dominate the early dynamics of massive discs, theoretical work suggests that this regime is relatively short-lived and difficult to observe \citep{hall2016, hall2019, rowther2020, rowther2022, rowther2023}, highlighting the need for accurate numerical simulations to improve our understanding of self-gravitating discs and their role in shaping planetary system architectures.\\

One key mechanism that can form planets or brown dwarfs, as well as impact the disc's future evolution, is disc fragmentation (\citealt{mb2010, vorobyov2010, stam2013, meru2015, schib2023, hans2025}). For fragmentation to occur in an infinitesimally thin disc, the stability parameter, $Q$, must fall below a critical value ($Q_{\rm crit} \approx1$) \citep{safronov, toomre1964}. $Q$ is dependent not only on the surface mass density, $\Sigma$, but also the sound speed, $c_{\rm s}$: 
\begin{equation}\label{eq:1}
Q = \frac{c_{s} \kappa }{\pi \Sigma G} ,
\end{equation}
\noindent
where $\kappa$ is the epicyclic frequency (which for Keplerian discs is equal to the angular frequency) and $G$ is the gravitational constant. Equation~\ref{eq:1} makes it apparent that for fragmentation to occur in a self-gravitating disc, either the disc needs to be sufficiently cool (low sound speeds) or enough material would need to accumulate onto a spiral arm (high surface mass density), or both.\\

\begin{figure*}
    \centering
    \includegraphics[width=\linewidth]{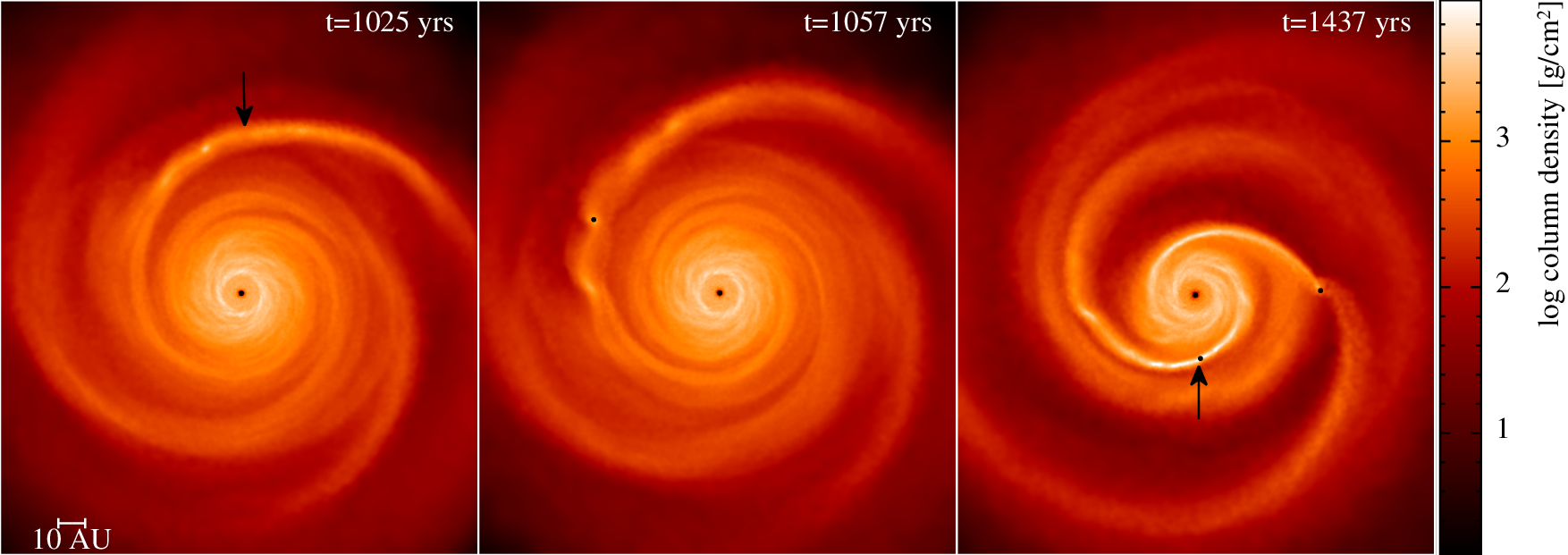}
    \caption{Surface mass density rendered images of \citealt{meru2015} Simulation 1 at 1.25~ORPs (1025 years), 1.30~ORPs (1057 years), and 1.76~ORPs (1437 years). Left to right: black arrow highlights the spiral arm that appears in the disc, following which the first fragment forms. This fragment triggers the formation of a second fragment (marked with a black arrow) at an inner location in the disc. Reproduced from \citealt{meru2015}.}
    \label{fig:splash}
\end{figure*}

\citealt{meru2015} used radiation hydrodynamical simulations to demonstrate that an initial fragment within a self-gravitating disc can trigger subsequent fragmentation, potentially leading to the formation of multiple bound objects. A key result from \citealt{meru2015} is reproduced in Figure~\ref{fig:splash}, where a second fragment has formed following the first - referred to as `Triggered Fragmentation'. This causal relationship, where early fragmentation induces further fragmentation, was attributed to changes in the surface mass density. \citet{meru2015} showed that the formation of the first fragment significantly enhanced the radial velocity of gas in parts of the disc, in some regions by up to a factor of $\approx$ 10. It was proposed that this increase in inwards transport of material raised the surface mass density in the inner disc, making the inner spirals dense enough to fragment further. However, Equation~\ref{eq:1} dictates that spiral overdensities are not the only factor controlling fragmentation and that the temperature can also impact it.\\

In this paper, we aim to address the question: Is the temperature or the surface mass density more important in driving self-gravitating structures to undergo triggered fragmentation? The answer has important implications for our understanding of disc fragmentation and, by extension, the composition and evolution of planetary systems that emerge from such environments. We reanalyse the simulations from \citealt{meru2015}, summarized in Section \ref{methods}. Section \ref{results} presents our key results, while Sections \ref{discussion} and \ref{conclusions} are dedicated to the discussion and conclusions, respectively.

\noindent
\begin{figure*}
    \centering
    \includegraphics[width=\linewidth]{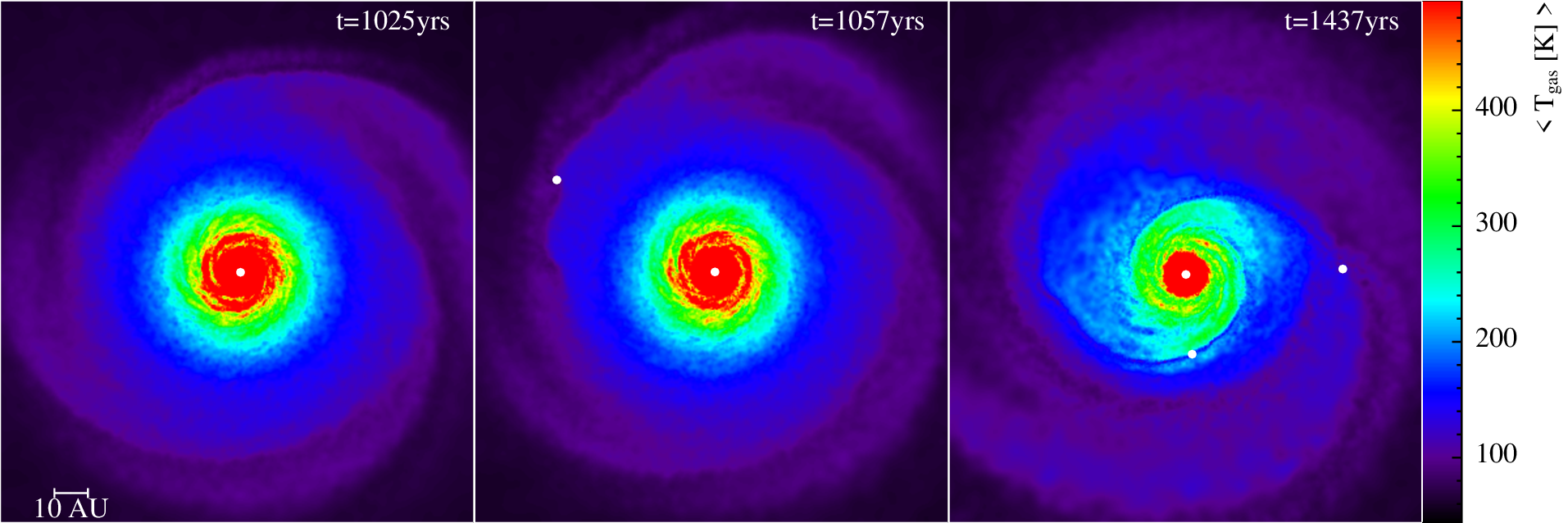}
    \caption{Gas temperature rendered images of Simulation 1 at 1.25~ORPs (1025 years), 1.30~ORPs (1057 years), and 1.76~ORPs (1437 years). Strong local temperature variations are seen at the location of the second fragment, following the formation of the first fragment. The fragments are marked by white filled circles.}
    \label{fig:temperature_2D}
\end{figure*}
\noindent

\section{Summary of Meru 2015: Triggered Fragmentation Numerical Setup and Results}\label{methods}

\citealt{meru2015} used three-dimensional smoothed particle hydrodynamics (SPH) simulations with radiative transfer to investigate the post-fragmentation dynamics of self-gravitating discs and the associated movement of gas in the disc. SPH is well-suited for modelling the non-linear dynamics of protoplanetary discs. Its Lagrangian nature allows it to effectively handle complex geometries and capture disc asymmetries \citep{fr2009, lp2010, price2012}.\\

Two discs with cylindrical radius $1 < R< 100$~au and a surface density profile $\Sigma \propto R^{-3/2}$ were simulated, both of which formed gravitational instabilities. The simulations encompassed two discs of different masses and stellar luminosities to explore fragmentation processes across different disc properties; Simulation 1: a  1.1\(M_\odot\) disc around a 1.5\(M_\odot\) star with luminosity \(L_\star\) = 4.3\(L_\odot\), and Simulation 2: a disc with a mass of 0.9\(M_\odot\) around a 1.0\(M_\odot\) star with \(L_\star\) = 2.4\(L_\odot\). Both simulations used 250000 gas particles.\\

The equation of state employed assumed a gas composition of hydrogen (70\%) and helium (28\%). This equation of state included considerations for the rotational and vibrational modes of molecular hydrogen, as well as the processes of molecular hydrogen dissociation and ionization of both hydrogen and helium. \citealt{meru2015} adopted a ratio of specific heats of 5/3 to characterize the thermodynamic behaviour of the gas. To account for the effects of stellar irradiation, they adopted a two-layer approach using the methodology followed by \citealt{mb2010}. Specifically, the mid-plane region was modelled using the flux-limited diffusion (FLD) approximation, which captured the transport of energy through the disc interior. The boundary layer of particles maintained a fixed temperature profile, which was set by the stellar radiation field. Any energy transferred to these boundary particles was effectively radiated away.\\

The particles at the interface between optically thick and optically thin regions were identified based on their vertical height. This is defined where the optical depth falls below unity. Here, we refer to this region as the disc `boundary'. In these simulations, the boundary height across a radial annulus was defined as the larger of two values: the height above the disc midplane where the optical depth was equal to unity \citep[see Appendix B,][]{meru2010}, and the height above which 10\% of the particles in the annulus were located. This dual condition ensures that there are enough boundary particles to model radiative cooling accurately, without allowing them to dominate the disc’s thermal evolution. Particles in this region are cooler than particles in the disc midplane due to the heating that results from the self-gravity. Grey Rosseland mean opacity values were adopted from \citealt{poll1985} for molecular dust grains, \citealt{alex1975} for gas contributions at high temperatures (for information on how the opacity tables were used, refer to \citealt{wb2006}). In order to investigate potential cooling effects due to interactions with the upper, optically thin regions of the disc, we follow this same prescription of \citealt{meru2015} in our reanalysis (see detailed SPH implementation in Appendix B of \citealt{meru2010}). \\

\citealt{meru2015} found that the radial velocity of gas within the disc significantly increased following initial fragmentation, with some regions experiencing up to a tenfold increase compared to pre-fragmentation velocities. This enhanced radial velocity was observed in both inward and outward directions, signifying substantial mass movement within the disc. Notably, the inward movement triggered by the initial fragment’s presence could lead to increased density in the inner regions, inducing further fragmentation at smaller radii than previously anticipated.\\

\citealt{meru2015} found that the first fragment in the $1.1\,M_\odot$ disc (Simulation 1) formed at $\approx54.5$~au at a time $t\approx1.30$ outer rotation periods (ORPs). This triggered the formation of the second fragment at $\approx24$~au at a time $t \approx1.76$~ORPs. For the $0.9\,M_\odot$ disc (Simulation 2), the first fragment formed at $\approx33$~au, at time $t \approx1.94$~ORPs. When this disc eventually fragmented again, the second fragment was formed at $\approx10$~au, at time $t \approx2.2$~ORPs. The rough time between the formation of the first and second fragments in both simulations was 300~years and initial masses of fragments ranged between 0.7 and 1.2~$M_\mathrm{J}$. To ensure that the formation of the second fragment was due to the formation of the first one, \citealt{meru2015} re-conducted the simulations but suppressed the formation of the first fragments in both simulations and allowed the discs to evolve. No subsequent fragmentation was observed in either of the simulations, demonstrating that the fragments formed in the initial calculations had been `triggered' by the initial fragment and that their radial location (between 10 and 25~au) was likely due to the inward movement of mass. These findings suggest that fragments may potentially form at smaller radii than previously anticipated by the disc fragmentation mechanisms \citep{rafikov2007, clarke2009, helled2014}.\\

\section{Results}\label{results}

\subsection{Tracing Instability: Radial Variations of $Q$, $c_s$, and $\Sigma$}

\begin{figure}
    \centering
    \begin{subfigure}[b]{0.85\columnwidth}
        \includegraphics[width=\linewidth]{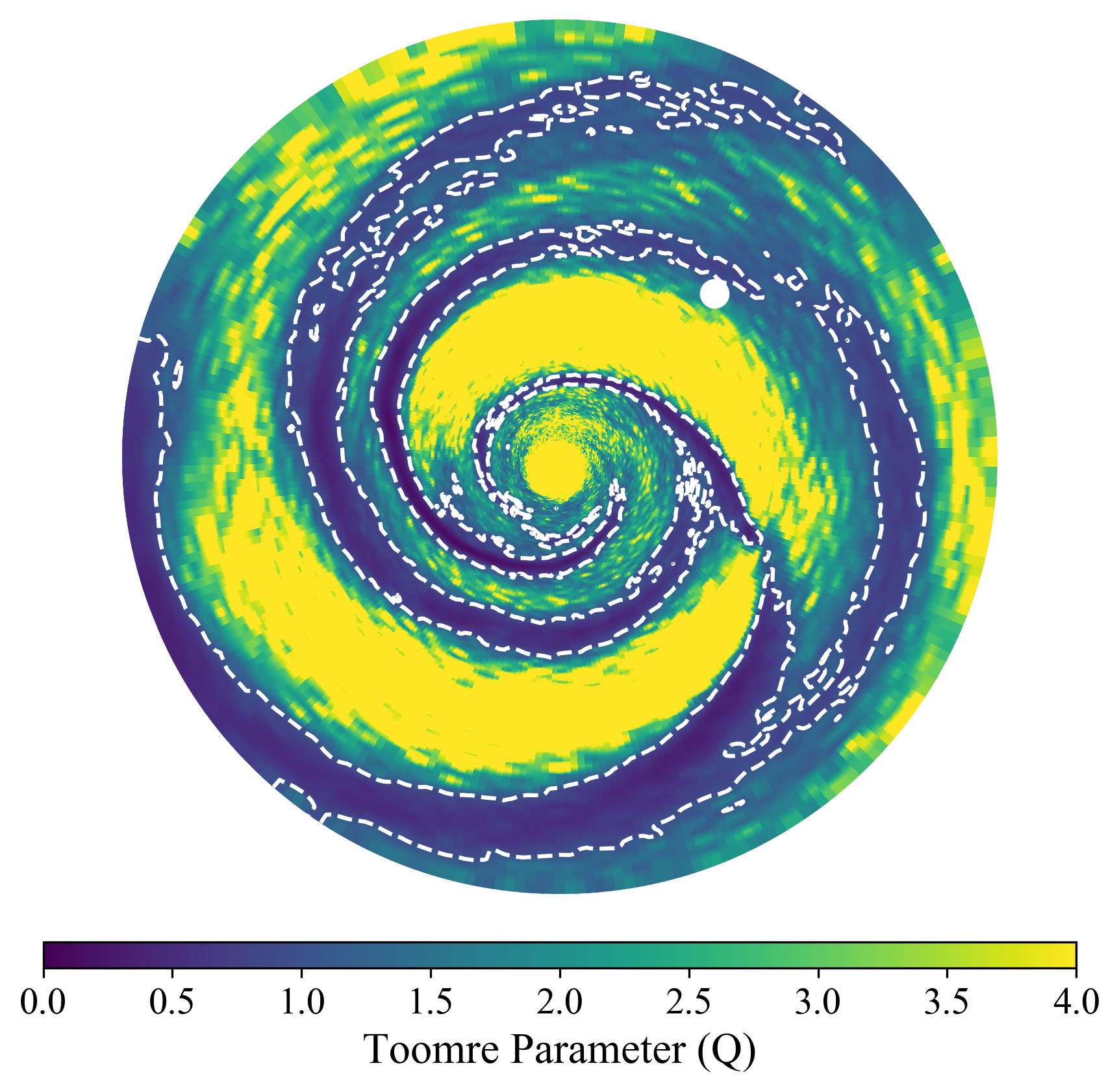}
        \caption{Simulation 1, $t = 1.76$~ORPs.}
        \label{fig:toomre_polar_sim1}
    \end{subfigure}
    
    \begin{subfigure}[b]{0.85\columnwidth}
        \includegraphics[width=\linewidth]{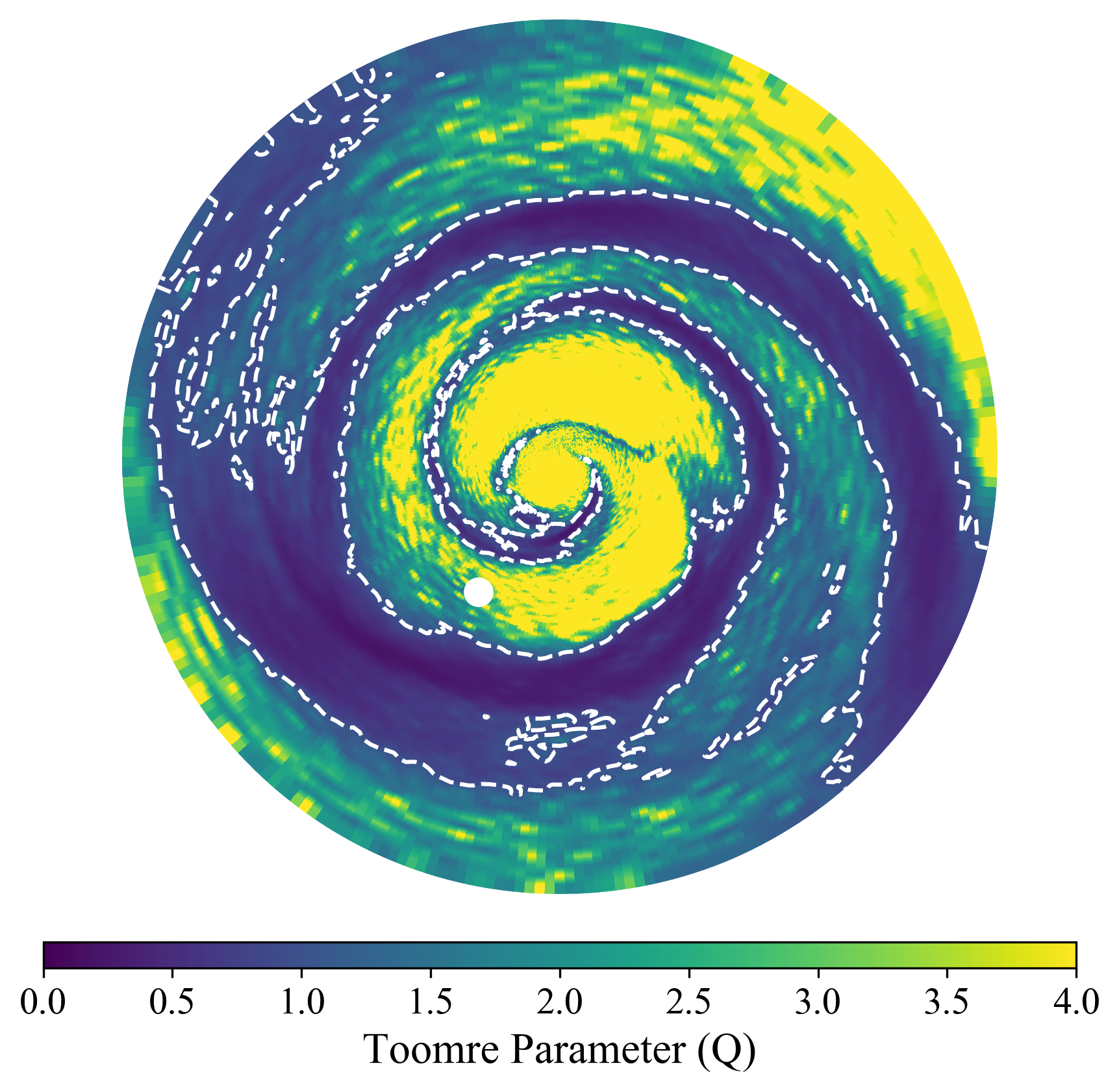}
        \caption{Simulation 2, $t = 2.24$~ORPs.}
        \label{fig:toomre_polar_sim2}
    \end{subfigure}
    
    \caption{2D maps of the Toomre stability parameter $Q$ for both simulations, shown immediately prior to the formation of the second fragment in each case. The colour scale ranges from gravitationally unstable ($Q < 1$, dark purple) to stable ($Q > 1$, yellow-green), with the dashed white contour marking the $Q = 1$ stability threshold. Transient spiral overdensities can locally satisfy $Q < 1$ without immediately fragmenting, as the disc goes in and out of the stability regime as it evolves. Where the instability $Q < 1$ persists though, is where fragmentation occurs \citep{gammie2001, rice2003, rafikov2005}. Outermost radius shown here is 100~au. The first fragment is marked by a white filled circle.}
    \label{fig:toomre_polar}
\end{figure}

To demonstrate what drives the disc to become gravitationally unstable and form a subsequent fragment, the azimuthally averaged density-weighted surface mass density, $\Sigma$, sound speed, $c_s$, and Toomre parameter, $Q$, are plotted for Simulations 1 and 2 in Figures~\ref{fig:radialprofilessim1} and \ref{fig:radialprofsim2}, respectively. While the value of $Q$ must be $<1$ for fragmentation to occur, we must take care when considering the azimuthally averaged $Q$. This is because certain localized regions of the disc satisfy the condition $Q<1$ and can fragment, while other regions may have higher $Q$ values. Azimuthal averaging can remove information about localized variations in $Q$, making it appear that fragmentation occurs in the disc even when the azimuthally averaged $Q$ at that radius is $>1$. In practice, local fluctuations in the surface mass density and sound speed (as seen in Figures~\ref{fig:splash} and \ref{fig:temperature_2D}) can still produce regions within a given radial slice where $Q < 1$. Figure \ref{fig:toomre_polar} demonstrates this effect with a 2D $Q$ plot for both simulations, showing the local variance in $Q$.\\

\noindent

\subsubsection{Simulation 1}
\begin{figure*}
    \centering
    \begin{minipage}{0.49\textwidth}
        \centering
        \includegraphics[width=\linewidth]{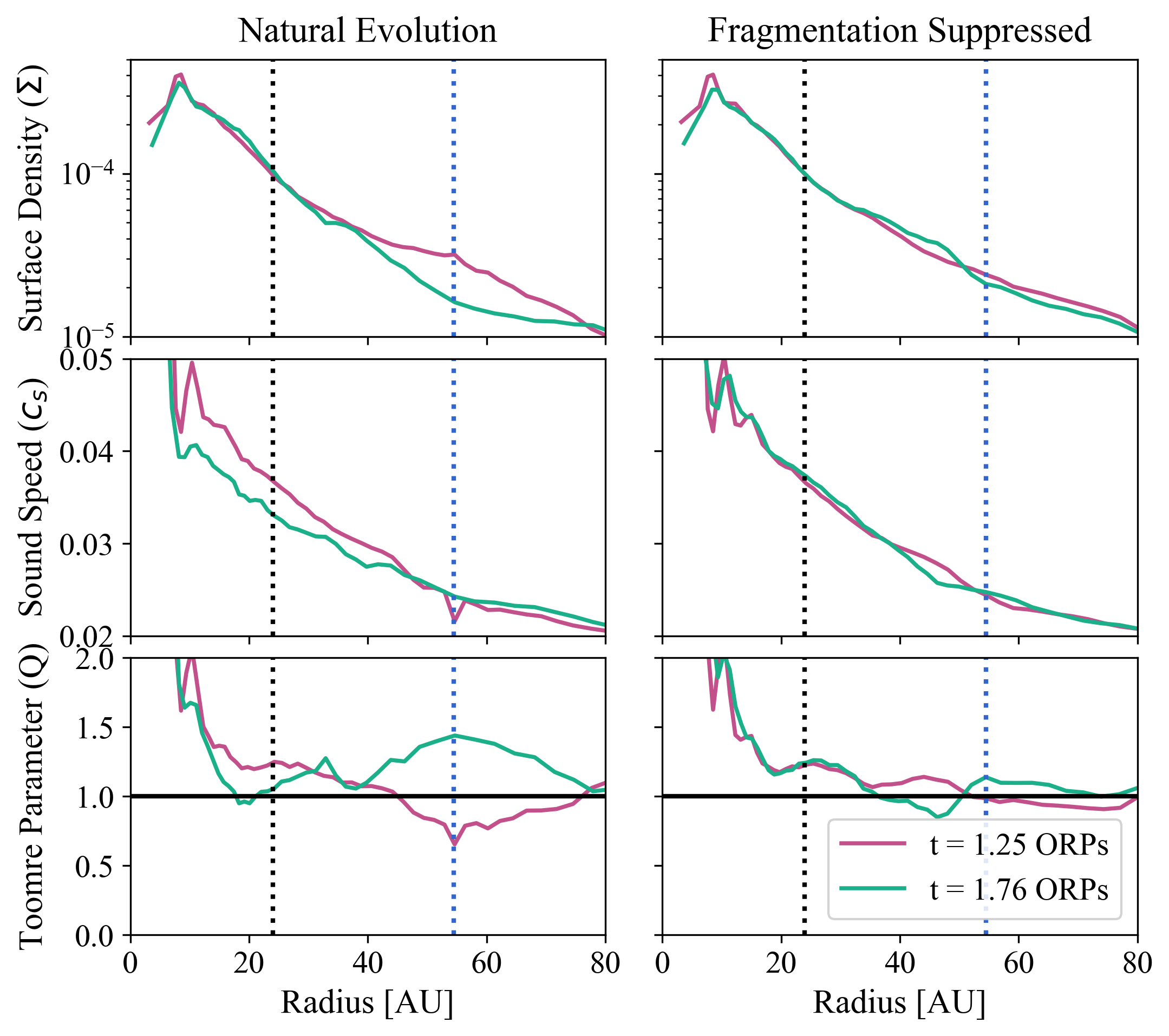}
        \caption{Simulation 1: Density-weighted, azimuthally averaged surface mass density, sound speed, and Toomre $Q$ parameter profiles at t = 1.25~ORPs (when the first fragment forms) to t = 1.76~ORPs (just before the formation of the second fragment). In the left panel, the blue and black dotted lines mark the radii where the first and second fragments form in Simulation~1. These radii are also indicated in the right panels for comparison, which show the corresponding profiles for the simulation in which the formation of the first fragment was suppressed and the disc did not fragment at all. The threshold for fragmentation, $Q = 1$, is marked with a solid black line. Both sound speed and surface mass density contribute to the decrease in $Q$ at the location of the second fragment (left). $Q$ remains higher than 1 at the location where the second fragment was supposed to form when the formation of the first fragment was suppressed (right). All quantities are in code units.}
        \label{fig:radialprofilessim1}
    \end{minipage}
    \hfill
    \begin{minipage}{0.49\textwidth}
        \centering
        \includegraphics[width=\linewidth]{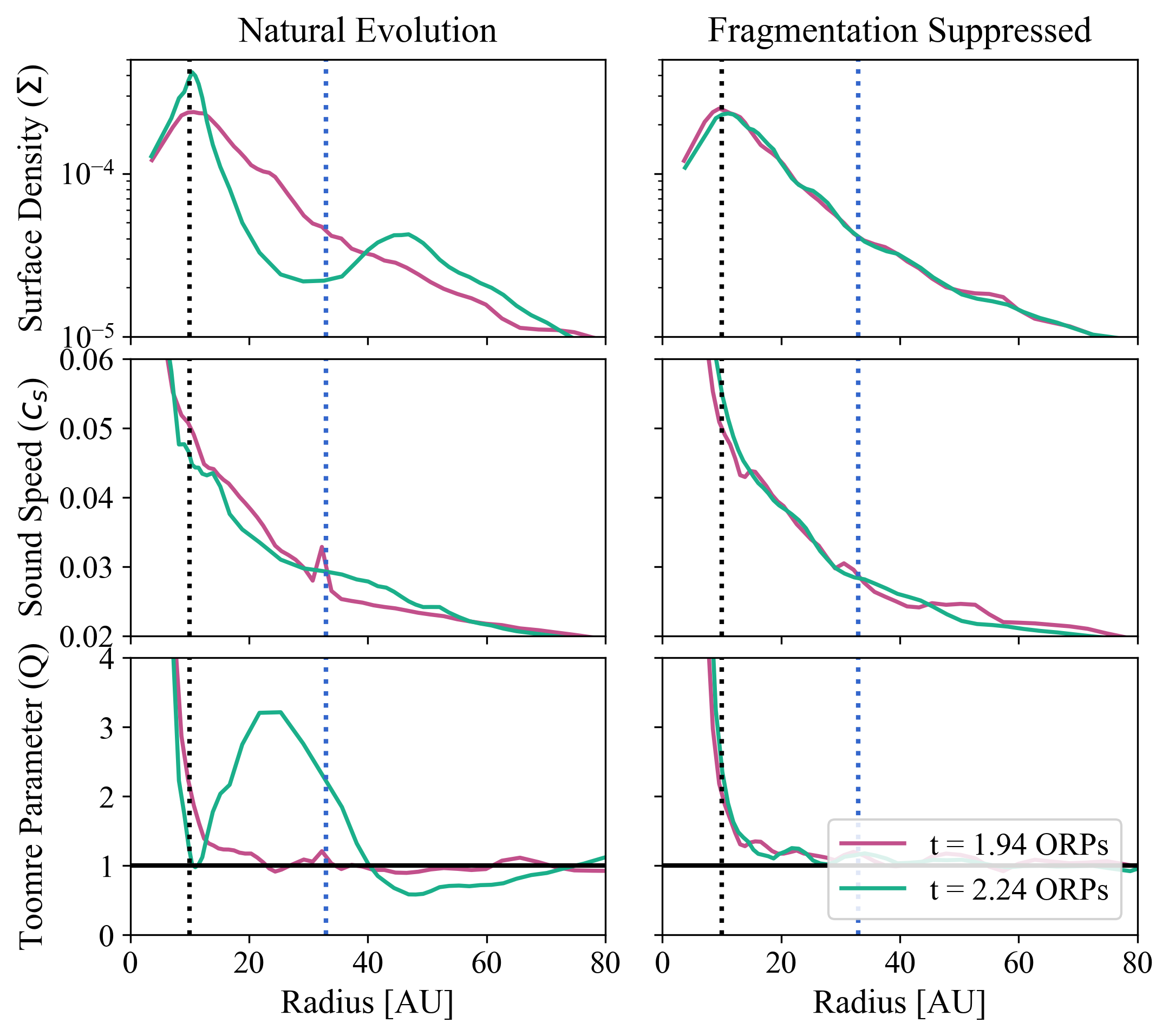}
        \caption{Simulation 2: Density-weighted, azimuthally averaged surface mass density, sound speed and Toomre parameter profiles from t = 1.94~ORPs (when the first fragment forms) to t = 2.24~ORPs (just before the formation of the second fragment). In the left panel, the blue and black dotted lines mark the radii where the first and second fragments form in Simulation~2. These radii are also indicated in the right panels for comparison, which show the corresponding profiles for the simulation in which the formation of the first fragment was suppressed and the disc did not fragment at all. The threshold for instability, $Q = 1$, is marked with a solid black line. Only surface mass density contributes to the decrease in $Q$ at the location where the second fragment formed (left). $Q$ remains higher than 1 at the location where the second fragment was supposed to form when the first fragment formation was suppressed (right). All quantities are in code units.}
        \label{fig:radialprofsim2}
    \end{minipage}
\end{figure*}

Figure~\ref{fig:radialprofilessim1} shows the density-weighted, azimuthally averaged surface mass density, sound speed, and Toomre parameter for Simulation 1, when the first fragment forms. The second fragment is located at 24~au where $Q$ decreases by $\approx$ 16\%, while $c_s$ decreases by $\approx$ 10\% and $\Sigma$ increases by $\approx$ 7\%. These changes show that both $c_s$ and $\Sigma$ play a roughly equal role in driving the reduction of $Q$, creating conditions that promote further fragmentation in the inner disc.  
This contrasts with the right panels, where the first fragment's formation was suppressed by artificially removing material from the disc. In this case, $Q$, $\Sigma$ and $c_s$ show negligible changes at the location where the fragment was supposed to form as the disc evolves and no fragmentation occurs.

\noindent

\subsubsection{Simulation 2}
For Simulation 2, Figure~\ref{fig:radialprofsim2} shows the $Q$, $c_s$, and $\Sigma$ as functions of radius. The left panels of Figure~\ref{fig:radialprofsim2}, which represent the setup where the first fragment forms show $Q$ decreasing by $\approx$ 23\%, $\Sigma$ increasing by $\approx$ 34\% and $c_s$ decreasing by $\approx$ 6\% at 10~au (the location of the second fragment). These results suggest that unlike Simulation 1, in Simulation 2 the reduction of $Q$ at 10~au is dominated by the increase in $\Sigma$, with $c_s$ playing only a minor role. In the right panels of Figure~\ref{fig:radialprofsim2}, which represent the case where the formation of the first fragment is suppressed and the second fragment does not form, the changes in surface mass density and sound speed are visibly smaller.

\noindent

\subsection{Surface Mass Density Evolution}

We extend the analysis done by \citet{meru2015} by examining the evolution of the surface mass density and verify that this inward mass movement contributes to the formation of the second fragment.\\

\begin{figure}
    \centering
    \begin{subfigure}[b]{\columnwidth}
        \centering
        \includegraphics[width=\textwidth]{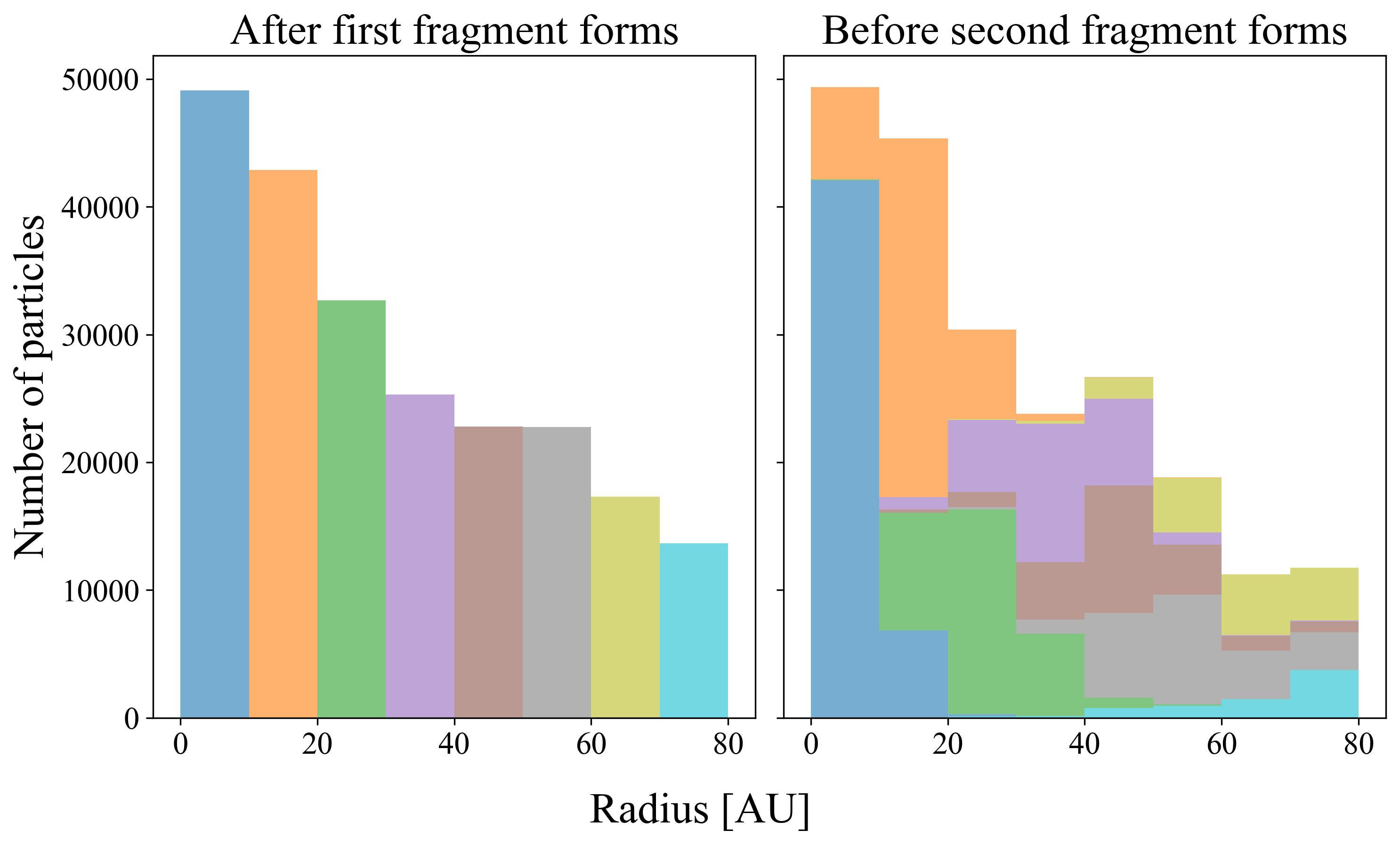}
        \label{fig:colorcoded_pair_sim1}
    \end{subfigure}
    \begin{subfigure}[b]{\columnwidth}
        \centering
        \includegraphics[width=\textwidth]{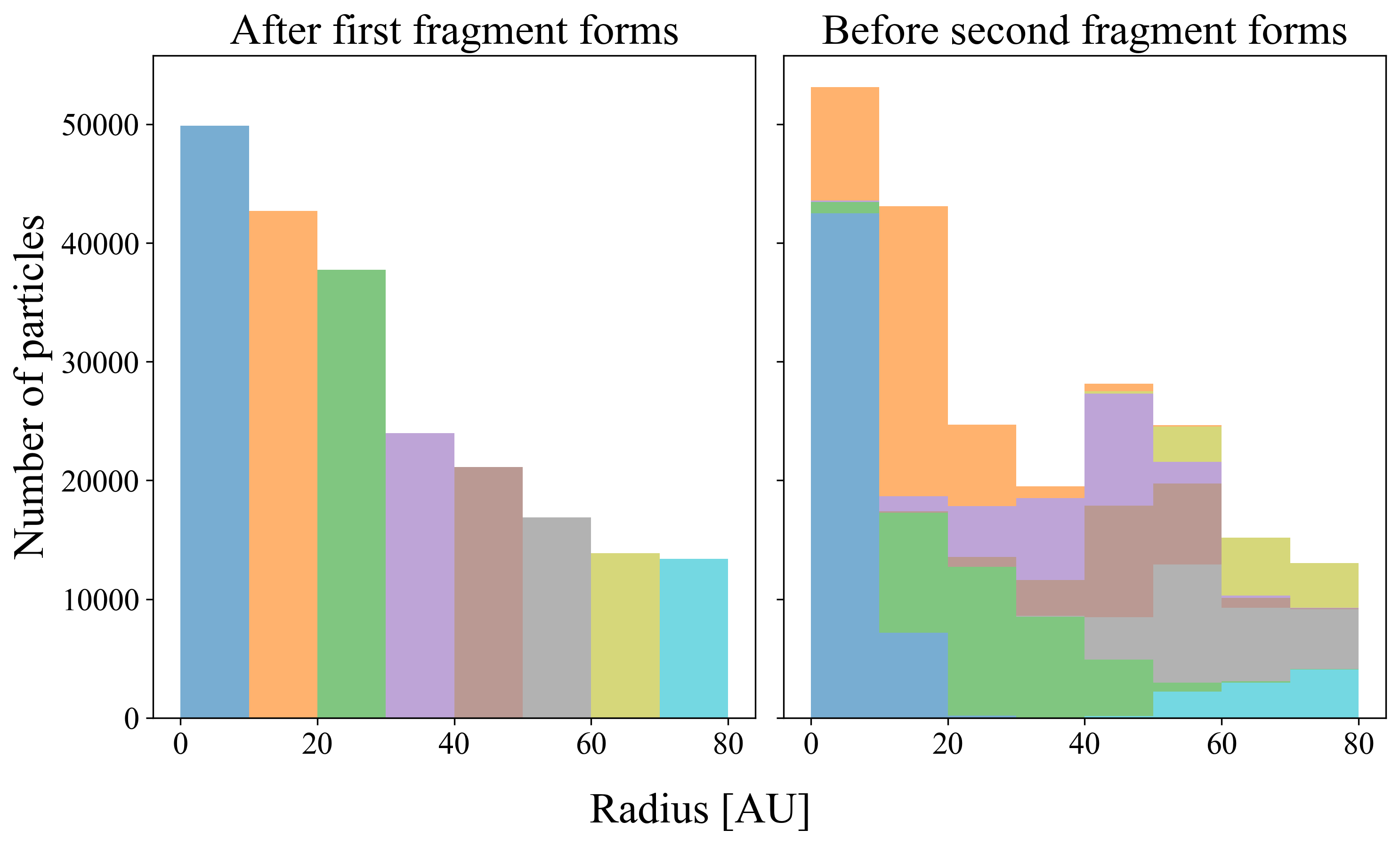}
        \label{fig:colorcoded_pair_sim2}
    \end{subfigure}
    \caption{
    Radial histogram of particles in the disc for Simulation 1 (top) and Simulation 2 (bottom), showing their distribution immediately after the formation of the first fragment (left) and shortly before the second fragment forms (right). Colours indicate the initial radial positions of particles at the time the first fragment forms. Each colour corresponds to a radial bin of 10~au, from 0–10~au up to 70–80~au. Strong mixing is seen near the second fragment radius in both simulations (24~au for Simulation 1, 10~au for Simulation 2), where material from the outer disc moves inward and material from the inner disc is scattered outward, consistent with the radial velocities reported in \citet{meru2015}. This redistribution corresponds to the increase in $\Sigma$ at the second fragment location and the decrease at larger radii, as shown in Figures~\ref{fig:radialprofilessim1} and~\ref{fig:radialprofsim2}, showing that initial fragment–driven mass transport enhances the surface mass density in the inner disc regions.
    }
    \label{fig:colorcoded_pair}
\end{figure}

Figures~\ref{fig:radialprofilessim1} and~\ref{fig:radialprofsim2} show the azimuthally averaged surface mass density profiles of the disc and how the mass is redistributed after the formation of the first fragment. At the radii where the second fragments form in both simulations, the surface mass density increases over time, while the region between the first and second fragments experiences a corresponding decrease. This supports the fact that material from larger radii is being redistributed inward, consistent with the mechanism suggested by \citet{meru2015}. Figure ~\ref{fig:colorcoded_pair} explores this further, distinguishing the particle locations and tracking their location. Figure~\ref{fig:colorcoded_pair} shows that in both simulations, some of the material located in the outer disc at the time the first fragment forms moves inward and enhances the surface mass density at the radius where the second fragment eventually forms.\\

Figures~\ref{fig:deltaTsim1b} and \ref{fig:deltaTsim2b} show the radial positions of all particles that eventually form the second fragment at the time of formation of the first fragment. These particles span a broad range of radii: between $\approx$ 5 and 80~au in Simulation 1 and between $\approx$ 5 and 40~au in Simulation 2, demonstrating that the second fragment is assembled from material drawn from across the disc, including particles that have moved inward from larger radii. Taken together, these results confirm that the enhancement of the surface mass density in the inner disc prior to the second fragment is driven by mass transport triggered by the first fragment.

\noindent

\subsection{Sound Speed Evolution}

To investigate what drives changes in $c_s$ at the location of the second fragment, we examine changes in the disc's temperature distribution driven by the formation of the first fragment. We define $\Delta T$ as the difference between the temperature of particles immediately before the formation of the second fragment and after the first fragment forms. This quantity captures any cooling or heating that occurs between the formation of the two fragments, which may contribute to triggered fragmentation.

\subsubsection{Simulation 1}
\begin{figure}
    \centering
    \begin{subfigure}[b]{\columnwidth}
        \includegraphics[width=\linewidth]{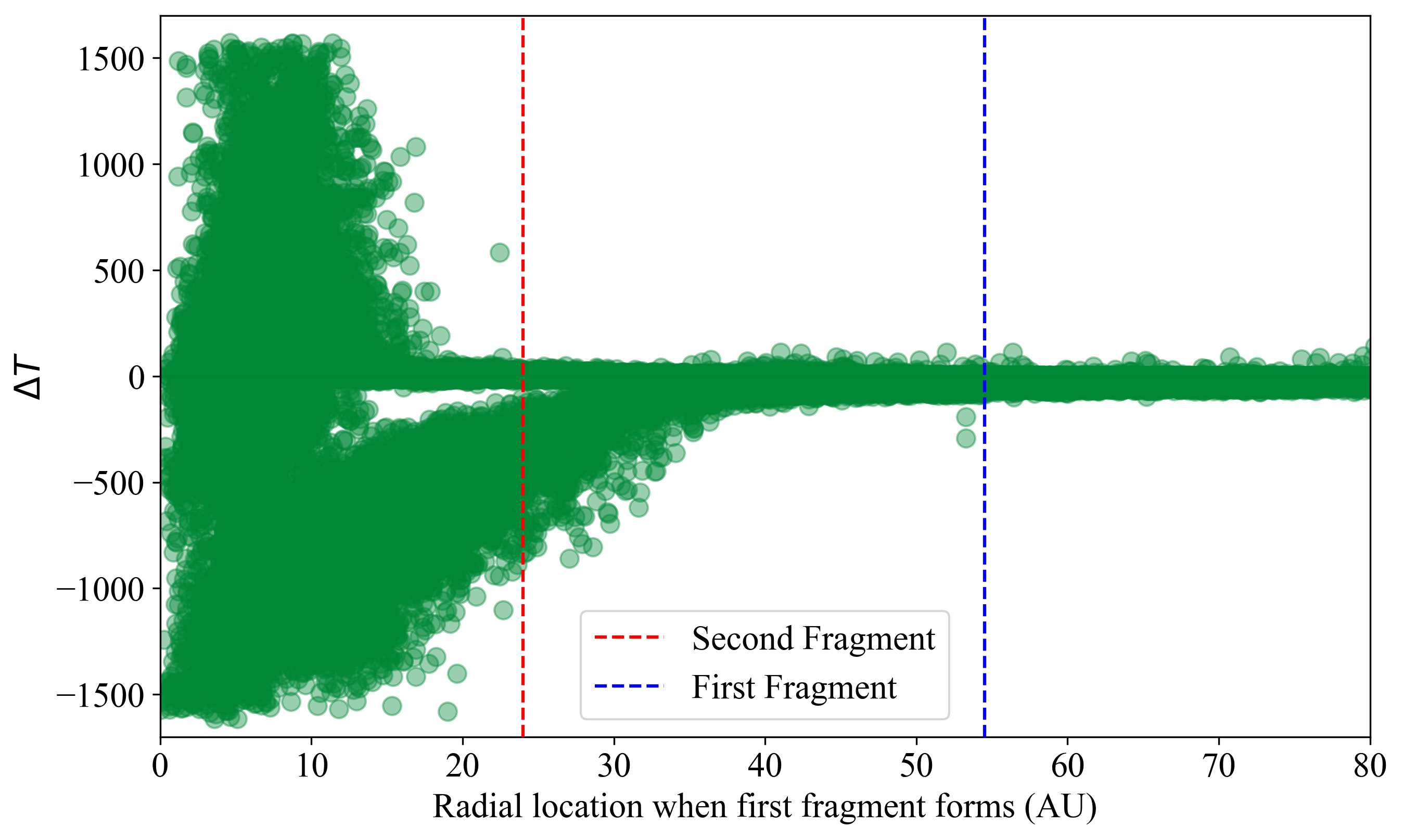}
        \caption{Simulation 1: $\Delta T$ versus $R$ for all particles in the disc. The formation locations of the first and second fragments are marked by dotted blue and red lines, respectively. At the location of the second fragment, the disc experiences cooling.}
        \label{fig:deltaTsim1a}
    \end{subfigure}
    
    \vspace{0.5cm}  
    
    \begin{subfigure}[b]{\columnwidth}
        \includegraphics[width=\linewidth]{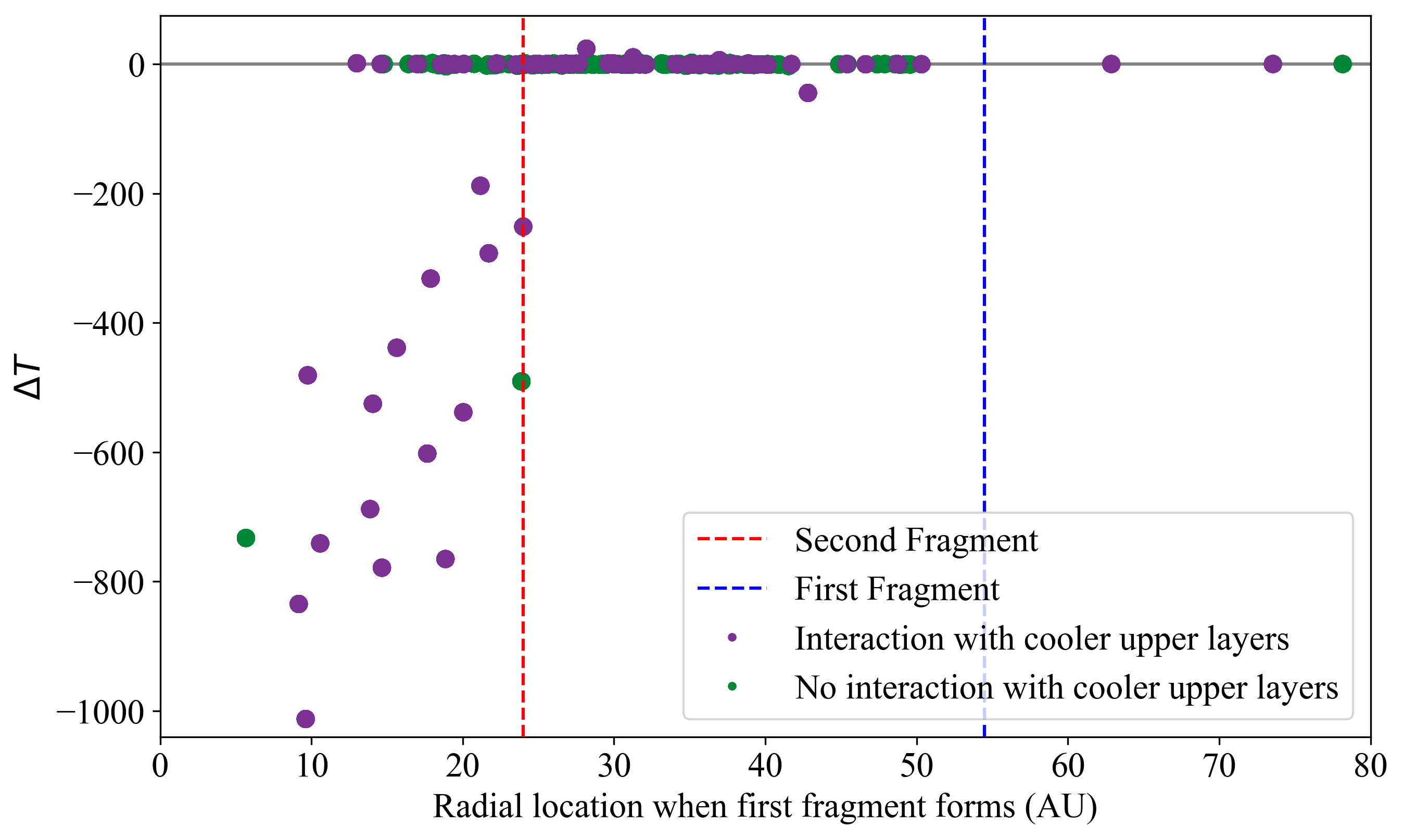}
        \caption{Simulation 1: $\Delta T$ versus $R$ for particles that form the second fragment. The particles that do not interact with the cooler disc surface in green, particles interacting with the cooler disc surface in purple. The majority of particles that end up in the second fragment either maintain their temperature or cool down.}
        \label{fig:deltaTsim1b}
    \end{subfigure}
    
    \caption{Simulation 1: Difference between the temperature of particles before the second fragment forms and after the first fragment forms, plotted against the radial location of particles just after the first fragment forms.}
    \label{fig:deltaTsim1}
\end{figure}

Figure~\ref{fig:deltaTsim1a} shows $\Delta T$ as a function of $R$ for all particles in the disc for Simulation 1, where $R$ represents the radial location of each particle shortly after the first fragment forms ($t = 1.30$ ORP). The figure shows that both cooling and heating occur across the disc; however, near 24~au --- where the second fragment forms --- most particles either maintain their temperature or cool down. Figure~\ref{fig:deltaTsim1b} shows $\Delta T$ versus $R$ specifically for the particles that go on to form the second fragment. The second fragment is composed largely of cooler material that moves inward from the outer disc while maintaining its temperature with a smaller contribution of material that travels outward from the inner disc and cools before it forms part of the fragment.\\

Approximately 90\% of all particles that form the second fragment maintain their temperature within a narrow range of $\pm 10$ K. Only $\approx$ 7\% experience significant cooling ($\Delta T < -10$ K), while the remaining fraction show heating ($\Delta T > 10$ K). The cooler material moving in from the outer disc has contributed to the inner disc becoming unstable and fragmenting. From Figure~\ref{fig:deltaTsim1b}, we also find that $\approx$ 50\% of the particles that constitute the second fragment have previously interacted with the disc boundary (i.e. cooler upper layers of the disc, as defined in Section \ref{methods}). Of these, $\approx$ 95\%  crossed the disc boundary at least once prior to fragment formation and concurrently showed a decrease in temperature, which also contributes to the observed decrease in the sound speed (see Discussion). This suggests that the sound speed reduces due to the movement of cooler material from the outer disc, driven by the formation of the first fragment but also due to interactions with the colder upper layers. 

\noindent

\subsubsection{Simulation 2}

\begin{figure}
    \centering
    \begin{subfigure}[b]{\columnwidth}
        \includegraphics[width=\textwidth]{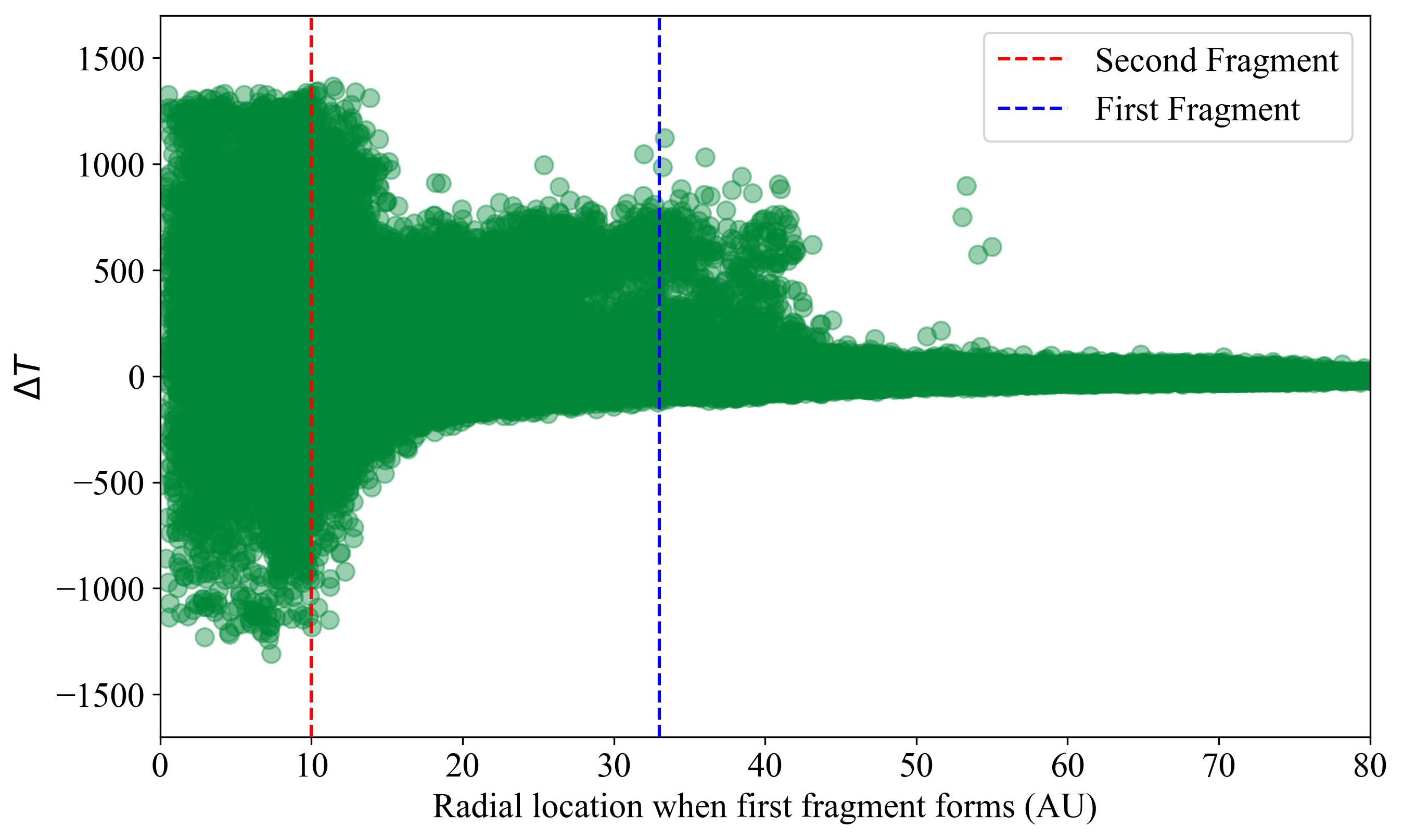}
        \caption{$\Delta T$ vs $R$ for all particles in the disc. The formation locations of the first and second fragments are marked by dotted blue and red lines, respectively. At the location of the second fragment, the disc experiences both heating and cooling.}
        \label{fig:deltaTsim2a}
    \end{subfigure}
    
    \vspace{0.5cm} 
    
    \begin{subfigure}[b]{\columnwidth}
        \includegraphics[width=\textwidth]{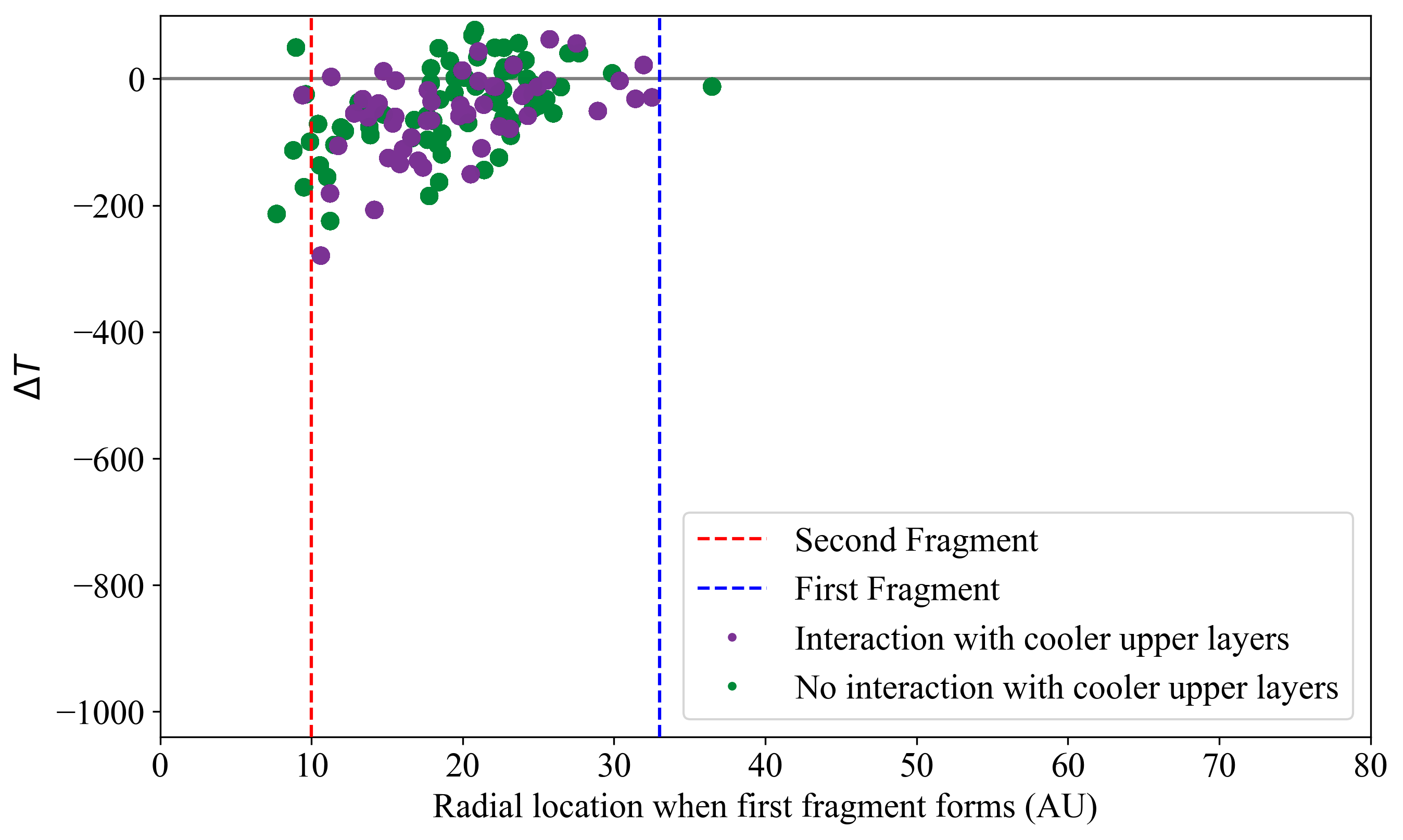}
        \caption{$\Delta T$ vs $R$ for particles that form the second fragment. The particles that do not interact with the cooler disc surface are in green, particles interacting with the cooler disc surface in purple. The majority of particles that end up in the second fragment either maintain their temperature or cool down.}
        \label{fig:deltaTsim2b}
    \end{subfigure}
    
    \caption{Simulation 2: Difference between the temperature of particles before the second fragment forms and after the first fragment forms, plotted against the radial location of particles just after the first fragment forms.}
    \label{fig:deltaTsim2}
\end{figure}

Figure~\ref{fig:deltaTsim2a} shows $\Delta T$ versus $R$ for all particles in the disc, where $R$ represents the radial location of particles right after the formation of the first fragment ($t = 1.94$ ORP). Both cooling and heating occur in the disc, even at 10~au --- the location where a second fragment eventually forms. Unlike Simulation 1, the $\Delta T$ distribution in Simulation 2 is more varied. Only $\approx$ 10\% of particles maintain their temperature within $\pm 10$~K, $\approx$ 30\% cool down, with $\Delta T < -10$~K, and $\approx$ 60\% experience heating with $\Delta T > 10$~K - indicating that a large fraction of particles heat up. \\

Figure~\ref{fig:deltaTsim2b} shows $\Delta T$ versus $R$ for only the particles that form the second fragment. The second fragment is made up of preferentially cooler material from the disc, selected out of a population that is both heating and cooling (Figure~\ref{fig:deltaTsim2a}). Approximately 70\% of the particles that make up the second fragment come from the cooled population, while only about 20\% of the heated particles end up contributing to the fragment.\\

Because the orbital timescale is shorter at 10~au (second fragment location in Simulation 2) than at 24~au (second fragment location in Simulation 1), the cooling (and heating) timescale is faster for the gas that forms the second fragment in Simulation 2 when compared to Simulation 1. While in Simulation 2 the particles have $\approx$ 250 years to heat or cool between the formation of the two fragments, the more efficient heating and cooling timescale allows for more efficient heating of the disc, which might explain the insignificant decrease in the azimuthally averaged $c_s$ in Figure~\ref{fig:radialprofsim2}.\\

The interaction of midplane material with the cooler, upper layers of the disc also has the potential to constrain additional heating at the formation site of the second fragment and facilitate its formation. Figure~\ref{fig:deltaTsim2b} shows that $\approx$ 50\% of the particles that contribute to the formation of the second fragment interact with the cooler upper layers (see Discussion). Of these, $\approx$ 93\% experience cooling upon interaction.

\noindent

\section{Discussion}\label{discussion}

We reanalyse the two simulations from \citealt{meru2015}. Our analysis demonstrates that in triggered fragmentation, \emph{both} $c_s$ and $\Sigma$ can play a role. In addition to the inward movement of material and hence increases in the surface mass density causing the inner disc to become unstable (as suggested by \citealt{meru2015}), the temperature can also contribute to the disc becoming unstable and fragmenting. Although the thermodynamic response is governed by local processes, such as compressional heating, radiative cooling, and changes in optical depth, these are influenced by the redistribution of material caused by fragmentation.\\

Triggered fragmentation provides a unique planet formation environment, as the inner fragment isn't just a product of its immediate surroundings; it is tied to the history of the first fragment and the material it ultimately pushes inwards. This type of dynamical mixing could redistribute material across the disc, suggesting a reconsideration of the common assumption of a radial chemical gradient in discs \citep{oberg}. Prior studies suggest that such physical processes that occur in the outer disc impact the chemical evolution in the inner disc \citep{ciesla2006, martin2012}, can redistribute complex organics \citep{ilee2021}, and lead to chemically distinct fragments - implications that extend to the diversity of forming planets and their atmospheres.\\

The mixing caused by triggered fragmentation may result in planetary architectures where neighbouring planets display starkly different compositions, or where widely separated planets in a system display similar compositions. This type of planetary architecture and its formation scenarios have been previously studied in theoretical work using core-accretion planet formation models and observations \citep{lokesh2023, lokesh2023b}, and our work identifies a formation pathway in which such architectures can also emerge from GI. Determining the core chemical composition of exoplanets remains challenging, as observations are limited to bulk density and atmospheric chemistry \citep{fortney2007, rogers2010}. However, triggered fragmentation-driven redistribution of material could explain and support the formation of planetary systems like TOI-125, Kepler-36, and WASP-47, where neighbouring planets show distinct compositions and widely separated planets occasionally share compositional similarities \citep{carter2012, lauren2017, nielsen2020}. Young systems such as L1448 IRS3B \citep{tobin2016, reynolds2021, rey2024}, where multiple compact sources appear to have formed in stages, may also offer a qualitative observational parallel to the triggered fragmentation scenario described here, though the complexity of such systems precludes a unique interpretation.\\

This redistribution of material also challenges current protoplanetary disc models used in planet formation simulations. Typically, these models assume a smooth initial dust distribution \citep{drka2023}, but dynamical mixing from triggered fragmentation could redistribute dust sizes across the disc before planet formation begins. This has significant implications for model outputs, as dust size distributions influence coagulation, settling, and planet formation efficiency \citep{testi2014}.\\

Due to the movement of mass caused by triggered fragmentation, inward transport of volatile-rich, cooler material becomes possible. This might result in dynamic thermal boundaries and potentially redistribute ices across classical snow surface locations \citep{ilee2017}. Although the canonical snowline radius is set by the local thermal balance (irradiation, viscous or GI heating, and radiative cooling), non-axisymmetric structures and fragments can locally alter both the temperature and the concentration of icy solids. Simulations of fragmenting, self-gravitating discs show that spiral shocks and fragment-driven heating produce multiple, spatially disconnected freeze-out fronts rather than a single radial boundary \citep{ilee2017}, supporting the idea that local structures can shift where ices appear or sublimate. Further GI models demonstrate that spirals and rings can generate several snowlines for the same species, driving substantial rearrangement of volatiles across condensation fronts \citep{tamara2025, vorobyov2023}.\\

In this context, fragments may sequester ices or be tidally disrupted and transport volatile-rich material inward, while spiral-driven radial drift can create local pile-ups or deficits across condensation fronts. These processes seen in triggered fragmentation can naturally produce local and transient shifts in volatile abundances, meaning that the apparent snowline location can vary even without any global change in the stellar irradiation field.\\

Early, GI-driven transport processes may thus play a key role in shaping the initial conditions for planetary compositions, which can differ from planets formed in-situ \citep{angelo2016, oberg2021}. Such chemical mixing also offers a plausible explanation for the partial chemical overlap seen between comets, asteroids, and planetary atmospheres \citep{bergin2007}. More broadly, this extends to the way we think about planet formation in general, as all GI discs will experience a dynamic GI phase leading to large-scale redistribution of dust and gas. This motivates the inclusion of chemical evolution modelling in future work and encourages further exploration of a disc's GI phase.\\

We found that the cooling process can be complemented by the interaction of disc midplane particles with the disc’s cooler, upper layers. By tracking the particles as a function of time, we find that these particles do not simply settle monotonically from the surface to the midplane, but instead undergo significant vertical circulation before being incorporated into the fragment, consistent with suggestions by \citealt{rl2018} that spiral structure in GI discs can drive vertical circulation. Additionally, while our simulations adopt the FLD approximation, which captures non-local radiative transport better than simple $\beta$-cooling, previous studies show that thermodynamic and fragmentation outcomes can depend on the radiative transfer treatment (differences between $\beta$-cooling, FLD and full radiative transfer are discussed in \citealt{kuiper2013, rowther2024}, and impact of the radiative modelling on disc fragmentation in e.g. \citealt{tsuka2015}). Recent simulations that couple hydrodynamics with live Monte-Carlo radiative transfer report differences in disc thermal regulation under irradiation: discs evolved with $\beta$-cooling reached a quasi-steady state where heating from spiral shocks balanced the cooling, whereas in irradiated discs with full radiative transfer, the stellar irradiation dominated the thermal evolution, stabilised the disc more quickly and reduced the longevity of strong gravitational instabilities \citep{rowther2024}. An important extension of the present work is therefore to verify whether the midplane–upper layer interactions and the associated cooling behaviour we identify within an FLD framework persist when full radiative-transfer methods are employed.\\

Future efforts to explore a broader range of disc and stellar configurations are also essential for capturing the complex interplay of physical processes that govern fragmentation in self-gravitating discs. Beyond variations in stellar and disc configurations, fragment luminosity and changes in disc opacity associated with fragment formation could also act to suppress fragmentation by modifying the local cooling and temperature structure of the disc \citep[e.g.][]{mb2010, nayakshin2013, mercer2017}. However, modelling self-consistent radiative feedback from bound fragments substantially increases the computational cost, and fragments are often treated using sink particles once they form \citep[e.g.][]{meru2015, nayakshin2026}. Assessing the impact of fragment feedback on triggered fragmentation is therefore left to future work.

\section{Conclusions}\label{conclusions}

In this study, we reanalyse the triggered fragmentation simulations from \citealt{meru2015}, which demonstrated that the formation of a first fragment in a gravitationally unstable disc can induce subsequent fragmentation.\\

We find that the formation of an initial fragment promotes further fragments at inner disc locations by impacting both the surface mass density and the temperature of the inner disc. The inward transport of material increases the surface mass density, while cooling can arise from both the inflow of cooler material and interactions with the disc’s cooler upper layers. Together, these effects demonstrate that variations in both sound speed and surface mass density can be expected to cause triggered fragmentation in GI discs.
\noindent

\section*{Acknowledgements}

We are grateful for the reviewers' thoughtful comments and suggestions on this manuscript. P.R. would like to thank Catherine Walsh, Cristiano Longarini, David Armstrong, and Dimitri Veras for interesting discussions. This research benefited from discussions at the Joint WG1 Meeting “Build it up”, P.R. acknowledges support towards participation from the COST Action CA22133 PLANETS. P.R. acknowledges support from a Royal Society Enhancement Award. F.M. acknowledges support from the Royal Society Dorothy Hodgkin Fellowship. R.N. acknowledges support from UKRI/EPSRC through a Stephen Hawking Fellowship (EP/T017287/1). R.N. acknowledges funding from the Australian Research Council via FT250100748. Computing facilities were provided by the Scientific Computing Research Technology Platform of the University of Warwick.

\section*{Data Availability}

The data underlying this article will be shared on reasonable request to the corresponding author. This work utilised the following public software:
\\
\textsc{Splash}: \href{https://github.com/danieljprice/splash}{https://github.com/danieljprice/splash} \citep{price2007}


\bibliographystyle{mnras}
\bibliography{example} 




\bsp	
\label{lastpage}
\end{document}